# Promises and Challenges of Ambient Assisted Living Systems

Hong Sun, Vincenzo De Florio, Ning Gui, Chris Blondia
*University of Antwerp*
*Department of Mathematics and Computer Science*
*Performance Analysis of Telecommunication Systems group*
*Middelheimlaan 1, 2020 Antwerp, Belgium, and*
*Interdisciplinary Institute for BroadBand Technology*
*Gaston Crommenlaan 8, 9050 Ghent-Ledeberg, Belgium*

## Abstract

*The population of elderly people keeps increasing rapidly, which becomes a predominant aspect of our societies. As such, solutions both efficacious and cost-effective need to be sought. Ambient Assisted Living (AAL) is a new approach which promises to address the needs from elderly people. Ambient Intelligence technologies are widely developed in this domain aiming to construct safe environments around assisted peoples and help them maintain independent living. However, there are still many fundamental issues in AAL that remain open. Most of the current efforts still do not fully express the power of human being, and the importance of social connections and social activities is less noticed. Our conjecture is that such features are fundamental prerequisites towards truly effective AAL services. This paper reviews the current status of researches on AAL, discusses the promises and possible advantages of AAL, and also indicates the challenges we must meet in order to develop practical and efficient AAL systems for elderly people. In this paper, we also propose an approach to construct effective home-care system for the elderly people.*

## 1. Introduction

As well known, the proportion of elderly people keeps increasing since the end of last century. The European overview report of Ambient Assisted Living (AAL) investigated this trend [1], and aims to find out an efficient solution to help these elderly people independently living.

Studies of Counsel and Care in UK found out that elderly people would prefer to live in their own home rather than in nursing houses, thus they need support to remain independent at their home [2]. Researches also proved that remote clinical therapy at home will not bring negative effect to the therapy process [3]. In order to improve the quality of life for the elderly people, it is important to guarantee that assistance to those people is timely arranged in case of need.

AAL aims at extending the time older people can live in their home environment by increasing their autonomy and assisting them in carrying out activities of daily living by the use of intelligent products and the provision of remote services including care services. Most efforts towards building ambient assisted living systems for the elderly people are based on developing pervasive devices and use Ambient Intelligence to integrate these devices together to construct a safety environment. Ambient intelligence refers to electronic systems that provide services in a sensitive and responsive way to the presences of people, and unobtrusively integrated into our daily environment [4] [5]. Living assistance systems and assistive devices are thus developed to facilitate the daily lives of these elderly people. These technologies promised to help the elderly people living independently in comfortable ways. However, their limitation is that these efforts still do not fully express the power of human being, and the importance of social connections and social activities is less noticed. Although such efforts are close to achieve the goal of assisting the elderly people to live independently by transferring the dependence from human side to assistive devices, we observe how such transfer also reduces the social connections of the assisted people.

In this paper, we take a broader view on AAL. We have seen the promise that home assistance systems developed using Ambient Intelligence could bring a safety environment around the elderly people. However, we also argue that the current solution overemphasized the importance of smart devices while either neglecting or lacking real implementations on the side of human interaction and human power. We will suggest combining the machinery power from assistive devices and the human power from social computing, seamlessly integrated together, timely provide needed services, and effectively utilize the social resources. This means not only to focus on keeping them physically healthy, but also to take their other daily requirements into consideration and best improve their quality of lives.

The rest of this paper is organized as follows: Section 2 reviews the work of building AAL systems for the elderly people. Section 3 discusses the main challenges and required works in developing AAL systems with human participation. Section 4 discusses a possible approach to construct effective AAL systems for the elderly people. Conclusions will be given in Section 5.

## 2. AAL Homecare Systems

Much research is being carried out on building intelligent environments around people, such as Aware Home [6] and I-Living [7]. These researches on “smart houses” improved the independence of the elderly people, and reduced the required manual work. Devices such as RFID, motion detectors, etc. are used to assist the daily lives of the elderly people. The Aware Home project built up a living lab, in which they tested the user’s acceptance of technology, building up a bridging framework for universal device interoperability in pervasive systems. Their researches also include discovering devices in pervasive computing environment, medical monitoring, and human computer interaction interfaces. The mission of I-Living is similar to that of Aware Home: developing an assisted-living supportive software infrastructure that allows disparate technologies, software components, and wireless devices to work together. Tasks provided in I-Living are such as activity reminding, health monitoring, personal belonging localization, emergency detection, and so on.

The above mentioned projects and many similar ones aim at providing assistive services in pervasive environments, construct a better environment and provide people with better lives. Services provided by those projects are promising to help the elderly people to ease their lives and keep them safe by monitoring some of their health status. However, the available services provided in those projects are still limited. The scenarios in these projects are still not complete enough to meet elderly people’s needs in their daily lives and help them maintain independent living.

The Amigo project [8], though not specifically designed for assisting the elderly people, investigated ambient intelligence for the networked home environment to provide attractive user services and improving end-user usability. Pervasive devices are managed in the Amigo project in an adaptive, context-aware and autonomous way. The system combines researches in home automation, consumer electronics, mobile communications and PC technology together to deliver services in a user centric way. The scenarios of this project proved that this system is able to provide the users with customized services. The applications are not restricted to the home environment, but extended to connect the work environment through mobile devices, and are also able to connect family members together.

The Amigo project is a huge step towards general introduction of the networked home and towards Ambient Intelligence by enlarging the usability of a networked home system. The achievements made in the Amigo project could be applied in Ambient Assisted Living for the elderly people to provide services by advanced ICT technology. However, as the Amigo project is not specifically designed to assist the elderly people living independently, there are challenges existing to fully express the potential of adopting those technologies to assist the elderly people independently living. Solutions of Amigo lack of human participation and the communication between the assisted people and the community out of their family, which inherently limits the service exploration, and may isolate the user from the outside world.

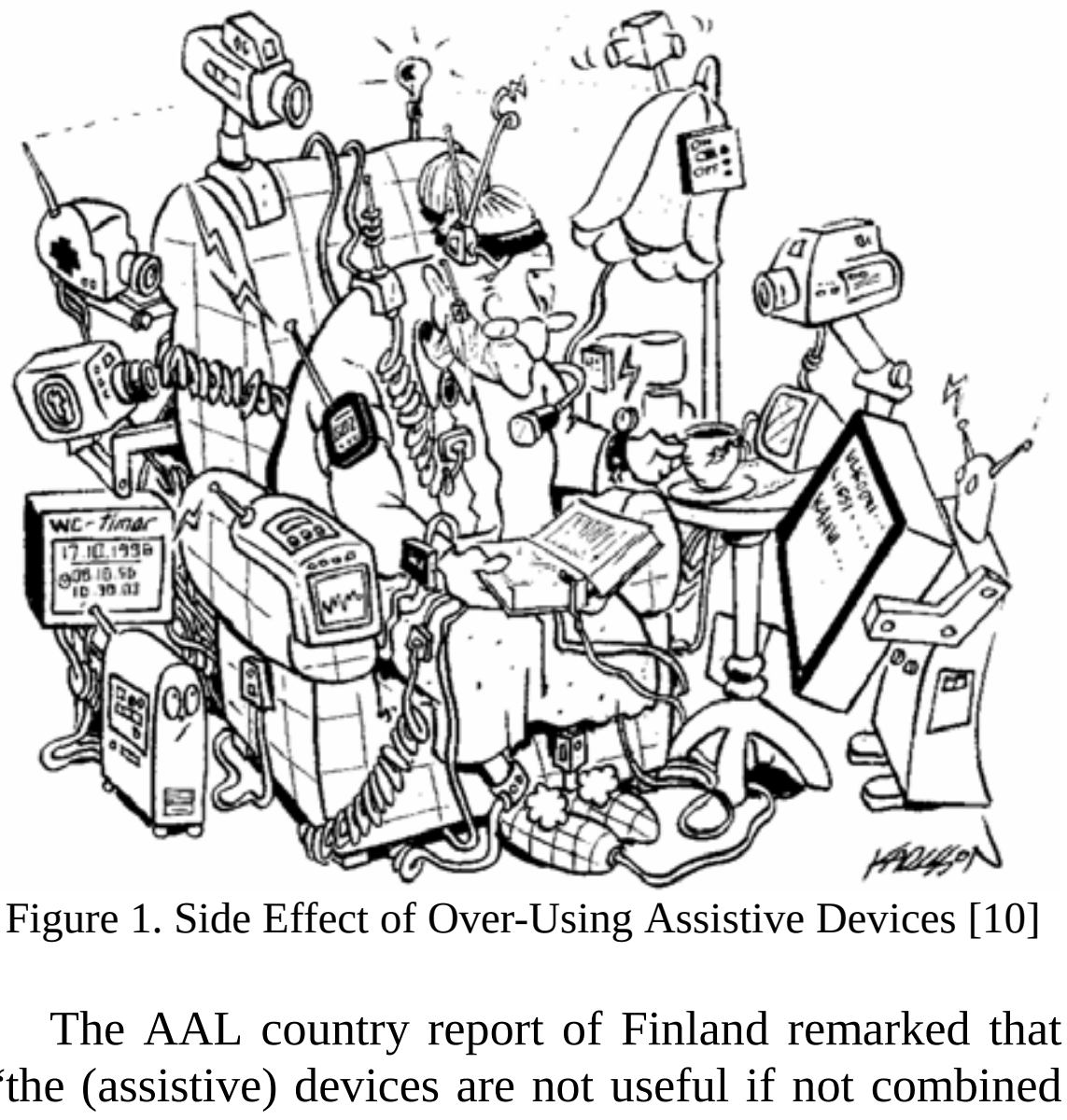

Figure 1. Side Effect of Over-Using Assistive Devices [10]

The AAL country report of Finland remarked that “the (assistive) devices are not useful if not combined with services and formal or informal support and help” [9]. We share this view and deem informal cares from relatives, friends and neighbouring people as indispensable when constructing timely and cost-effectively services to assist the elderly people. The usage of assistive devices helps transfer the dependence from human side to machinery side, thus establishing some degree of independence. However, the dependence on the assistive devices unconsciously reduces the social connections of the assisted people. Without the communications with the outside world, elderly people assisted by those assistive devices are only safely surviving, but not actively living. Figure 1 shows the possible side-effect of over-using technology without proper human participation. Although the effect is exaggerated, the picture reminds us that we should be cautious not to leave the elderly people only with assistive devices, but also with our compassions and communications.

Much effort is also being made to connect the assisted people together, and carry out services based on communication between human beings. One such project is COPLINTHO [11], which built an eHomeCare system combing forces from the patient's family, friends and overall care team. The limitation of this class of investigations is that the application is specialized on the recovery progress of a patient, thus the communications are mainly focused on exchanging the medical data of the patient, which restricted more generalized application.

All the above are examples showing current efforts to assist the elderly people living independently in their own houses. We have already seen that the advanced Ambient Intelligence technology is able to construct safety environment around the elderly people, provide them with customized services, and improve their quality of lives. We also observed that there are projects that intend to address the elderly people's needs by connecting them together, increasing their communications and keeping them active in society. However, we also find that the above mentioned solutions are either focused on technological aspect or societal aspect. We argue that effective and efficient solutions to meet the AAL challenges should combine the forces from both the technological part and the societal ones. The participations of human beings could help fully express the potential of smart devices, and maintain the social awareness of the elderly people; the usage of advanced ICT technology could better connect the elderly people together, organize community activities. In the following session, we firstly discuss the challenges of developing pervasive home-care services with human participation, and then we propose a possible approach to construct such a system.

## 3. Challenges of AAL Systems

The previous sections stated the promising approaches to help the elderly people living safely and independently in their own houses. We also proposed to integrate the human services into AAL systems to enrich the available services and create a less intrusive environment. However, there are still many challenges towards the implementation of such an environment. In this section, we will discuss those we deem to be the main challenges, the necessary steps, and some possible solutions to effectively deliver services in AAL system and bring in human participation.

### *Challenge 1. Dynamic of service availability*

Although informal caregivers may help reduce the needed social resources, and increase the social connections, they are also very difficult to be utilized, and inherently very dynamic: the availabilities of these services are continuously changing. How to manage this dynamicity becomes a big challenge.

Service Oriented Architecture (SOA) could be a good approach to cope with the above mentioned dynamicity. SOA is a flexible, standardized architecture that supports the connection of various services, and as such is an ideal tool to tackle the dynamicity problem. The application of SOA, such as the OSGi platform [12], can also help to establish a framework such that various smart devices could be integrated together and could be automatically called, started or stopped. The regulated service format will also help the process of service matching. Research efforts of using OSGi to build safety home environment are also reported in [13].

Another attractive feature of SOA can be found in recent researches towards integrating people activities into service frameworks, which culminated in two specifications launched in the summer of 2007: WS-BPEL Extension for People (BPEL4People) [14] and Web Services Human Task (WS-HumanTask) [15]. The WS-HumanTask targets on the integration of human beings in service oriented applications. It provides a notation, state diagram and API for human tasks, as well as a coordination protocol that allows interaction with human tasks in a service-oriented fashion and at the same time controls tasks' autonomy. A people activity (service) could be described as human tasks in the WS-HumanTask specification. The BPEL4People specification supports a broad range of scenarios that involve people within business processes, using human tasks defined in the WS-HumanTask specification. These two specifications could help to meet the challenges of integrating human services in the SOA framework of the proposed mutual assistance community.

*Challenge 2. Service Mapping*

How to let the computer automatically map the available/requested services is a big challenge in AAL system.

The foundation for service mapping is service description. A Semantic Knowledge Base is required to precisely describe the advertised services: certain ontology libraries describing the domain knowledge of the home-care environment should be developed with the collaboration of the researchers in this domain. With such domain knowledge, conceptual model for semantic service matching could be applied. OWL-S [16] is currently the most used technology in this domain; it is able to provide a framework for semantically describing web services from several perspectives, for instance, service inquiry, invocation, composition.

There are some service matching tools developed for matching OWL-S services, such as OWL-S Matcher [17], OWL-S UDDI/Matchmaker [18], and OWLS-MX Matchmaker [19]. The drawback of the first two is that the matching process takes a large amount of time, while the drawback of the latter one is being memory intensive [20]. These tools serve as good starting points to investigate web service matching, while we believe more elegant and efficient matching engines should be developed. We have made some preliminary tests of service matching in home-care service matching – details can be found in our paper [21].

*Challenge 3. People's Willingness*

People's willingness to participate in AAL systems needs to be investigated and encouraged – how to encourage people joining e.g. a mutual assistance community (see Section 4) is a big challenge.

We deem that any truly effective AAL system cannot leave aside the contributions coming from society itself, in all forms, with the participation of informal caregivers, professionals, and even the elderly people themselves. In order to encourage more people to make contribution to AAL system, we need to understand their drives to provide help to others, and stay active in the community. The main drive for people to help others is not merely money, but also includes moral duty and their social image. One main reason that keeps people active in an online community is to build up a good image for their avatars and win respect from other community residents. An AAL system with participation of informal care-givers could also reward the informal caregiver in this way. Social studies to stimulate people to work as volunteers should be thoroughly carried out.

Besides the willingness to help others in the mutual assistance community, elderly people's willingness to make use of their assisting systems also needs to be studied and encouraged. We would ascribe the elderly people's unwillingness to use assisting system from two folds – psychologically and technologically. In the rest of this section we introduce these two aspects in details and explain how we made our efforts to reduce this unwillingness

*Psychological Frustration:*

When people are getting old, a relevant source of frustration comes from losing physical strength, but what torches them most lies in psychology: they are becoming passive consumers of the societal services rather than active creators. In so doing, they also lose their self-esteem. Almost all the AAL systems for the elderly people consider their users as people who are weak and passively assisted by others. For the designers of such systems, being able to maintain some degree of independence without bringing too much burden to our society appears as an already ambitious goal. However, those systems neglect the fact that the elderly people can still make their contributions to our society through their valuable experiences. A home-care system with human participation could help encourage the elderly people to actively participate in group activities as peer participants, and possibly even to use their experiences to help the younger generations solve, e.g., some of their work and school problems [22]. Such possibilities will be discussed in more details in the following section.

*Technological Frustration:*

Elderly people are usually scared by the application of new technology. In order to help them get used to the ambient assistive devices, we should construct user friendly interfaces, and also provide appropriate trainings to their users. Developing adaptive, natural and multimodal human computer interfaces is the main challenge of future interfaces in assisted living [23]. It is also suggested to get people involved in how to use the assistive devices before they really need them [24].

In the solution proposed in next section, elderly people do not only benefit from keeping connected with other people, but also are provided chances to make contributions to society, to feel that they are still useful, and to live in an active way with self-esteem. Such benefits could provide our elderly people with stimulations to break the technical barrier.

## 4. Mutual Assistance Community - A Promising Approach

The previous section discussed the challenges of developing AAL systems with human participation. In this section, we propose to construct a so-called "mutual assistance community" to bring services from human side into AAL environments. Such a system integrates the services from human beings with the applications provided by the devices, and best utilizes the available resources and provides services to the people in need in an effective way [25] [26] [27]. Figure 2 shows the organization of our mutual assistance community. Assistive devices will be deployed to construct a smart house environment managed by a local coordinator to build up a safety environment around the assisted people.

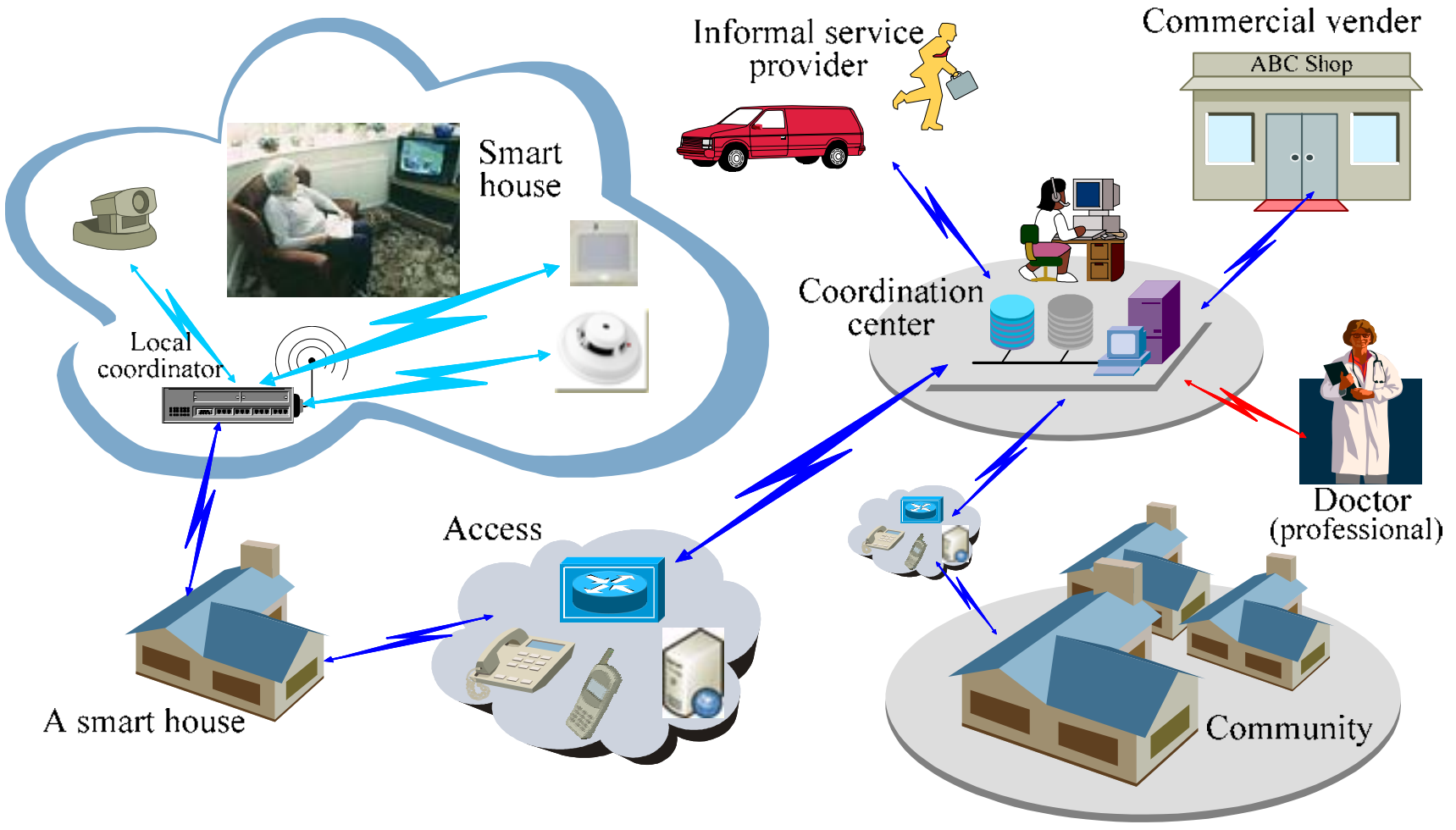

Figure 2. Organization of Mutual Assistance Community

The most important asset integrated in our community is indeed the people themselves. Our proposed community allows disparate technologies and people working together to help people who suffer from aging or disabilities. People who are able to provide services are encouraged to do so and assist the requesting people as informal caregivers. Elderly people are also encouraged to participate in the group activities, which not only helps to maintain physical and psychological health but also reduces the requests of professional medical resources. Professional caregivers (such as doctors, specialists etc.) are included in the community to provide emergency and professional medical service. Commercial vendors are also included, which brings convenience to the user and diversifies the service type, at the same time laying the foundation for economical exploitation and self-sustainability.

In fact, the importance of the informal caregiver is also notified by the projects reviewed in this paper: for instance, Aware Home points out that technology should support networks of formal and informal caregivers, and the scenario of Amigo project also shows informal caregivers could help to provide first-aid help to neighbouring people. However, the link between the informal caregivers and the elderly people are statically fixed there. Our proposed community can flexibly connect the needed help and available informal caregiver services through web service publication, matching and binding. The elderly people can also use this approach to initialize and join group activities, and inter-generational activities could also be carried out in this way. During the inter-generational activities, the younger generation could help the elderly people on physical strength demanding tasks as informal caregivers. Though physically weak, the elderly people accumulated valuable experiences and knowledge during their lives. They may use such knowledge to assist the younger generation in solving their problems in works and studies. During this process, not only the younger generation gets their needed answer, the elder generation also finds an access to make their contribution to our society. The elderly people may find themselves still useful, stand with more active living attitude, thus avoiding the frustration of being useless.

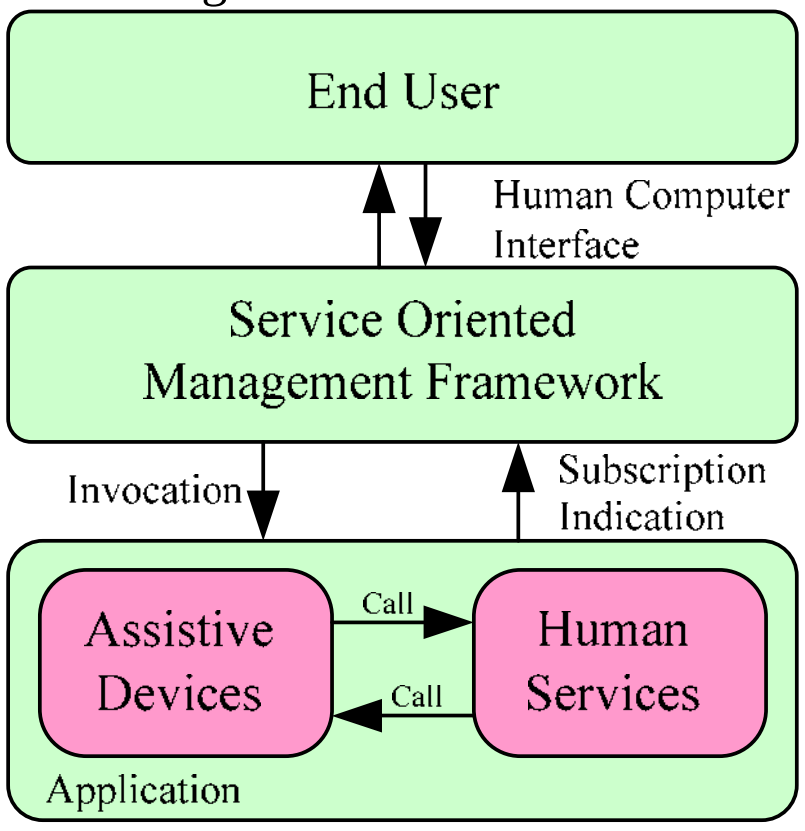

Figure 3. Service Organization

The interactions between the assistive devices, human services, and end users, under the coordination of a service oriented management layer, are shown in Fig. 3. Assistive devices and human services will interactively work together to express potentials from both sides providing high quality services to the people in need. Both assistive devices and human services will subscribe their availability through a service oriented management framework. Requested and available services will be reasoned in the framework, and matched available service will be invocated. End users will interact with the framework through human computer interface. In the proposed mutual assistance community, a user could be registered as a service consumer and a service provider at the same time, based on the type of services. For instance, an elderly person could request physical challenging services, while providing knowledge/experience-based advisory/consulting services.

Simulation results demonstrate that the presence of informal caregivers helps reduce the social resources and provide daily assistance timely [25]; and when the elderly people are actively participating group activities, the dependence on social resources are further reduced, while their social connections are kept and even strengthened [26].

## 5. Conclusions

This paper discussed the current issues of building AAL systems for elderly people. We have observed the research efforts of building pervasive home-care environments with advanced Ambient Intelligence, which promises to provide safety environments around the elderly people in their own houses. However, we also foresee some challenges of such solutions: we consider a concrete threat the possible social isolation due to the over-use of technology and lack of the communication between the assisted people and the outside community. We believe effective and efficient solutions to assist the elderly people independently and actively living should leverage the efforts from both technical side and social side. We are very glad to see that now there are more efforts from the government side focusing on the social connections between the assisted people and the outside world, as witnessed e.g. by the second call of the European AAL programme, which stated the importance of helping the elderly people live actively and enjoy their life, bridging distances and preventing loneliness and isolation [28].

We deem that human participation could help to meet the challenges of building an efficient and effective AAL system. Indeed, human participation makes it possible to alleviate the sense of intrusion, to diversify the service categories, to explore the potential of the assistive devices, and also provides the elderly people with chances to serve our society and keep living actively.

Mutual assistance community, where people mutually assist each other, is recommended as a possible approach to bring human services into AAL systems. Smart devices can still be used in such a community to guarantee the safety of elderly people. Elderly people can also actively maintain their social networks, and regain their self-esteem. We are convinced that bringing the people in home-care environment is both efficacious and effective in saving the social resources, providing timely services and helping the elderly people live in an active way.

## References:

[1] H. Steg, et al. Ambient Assisted Living – European overview report, September, 2005.

[2] Counsel and Care, Community Care Assessment and Services, April, 2005.

[3] J. Deutsch, J. Lewis, and G. Burdea, Technical and patient Performance with a Virtual Reality-Integrated Telerehabilitation System, I*EEE Transactions on Neural Systems and Rehabilitation Engineering, Vol. 15(1), pp. 30-35*, March 2007.

[4] E. Aarts, R. Harwig and M. Schuurmans, chapter Ambient Intelligence in The Invisible Future: The Seamless Integration Of Technology Into Everyday Life, *McGraw-Hil*, 2001.

[5] E. Aarts and J. Encarnação, True Visions: The Emergence of Ambient Intelligence, *Springer*, 2006.

[6] Aware Home. Aware Home, Georgia Institute of Technology, http://www.cc.gatech.edu/fce/ahri/, 2008.

[7] I-Living. University of Illinois at Urbana-Champaign, Assisted Living Project. http://lion.cs.uiuc.edu/assistedliving

[8] Amigo – Ambient Intelligence for the Networked Home Environment. http://www.hitech-projects.com/euprojects/amigo

[9] Ambient Assisted Living, country report, Finland, 2005.

[10] Risto Karlsson *Printed in Helsingin Sanomat 18.10.* 1996.

[11] COPLINTHO. COPLINTHO, IBBT, Innovative Communication Platform for Interactive eHomeCare. https://projects.ibbt.be/coplintho/, 2008.

[12] Open Service Gateway initiative (OSGI), OSGI Service Platform Version 4, http://www.osgi.org, 2008.

[13] M. Aiello, S. Dustdar. Are our homes ready for services? A domotic infrastructure based on the Web service stack. Pervasive and Mobile Computing 4(4).2008.

[14] Specification: WS-BPEL Extension for People, (BPEL4People), version 1.0. 2007.

[15] Specification: Web Services for Human Task (WS-HumanTask), version 1.0. 2007.

[16] OWL-S Technical Committee (T.C), Web Ontology Language for Web Services (OWL-S), http://www. w3.org/Submission/OWL-S/, 2002.

[17] S. Tang. The TUB OWL-S Matcher, http://owlsm.projects.semwebcentral.org/, 2008.

[18] Naveen Srinivasan. OWL-S UDDI Matchmaker. http://projects.semwebcentral.org/projects/owl-s-uddi-mm/, 2008.

[19] M. Klusch, B. Fries, M Khalid, K Sycara. OWL-MX Matcher. http://projects.semwebcentral.org/frs/?group_id=90, 2008.

[20] N. Georgantas,et al. Amigo Middleware Core: Prototype Implementation & Documentation. IST Amigo Project Deliverable D3.2, 2006.

[21] H. Sun, V. De Florio, N. Gui and C. Blondia. Service Matching in Online Community for Mutual Assisted Living. *In the Proceedings of The Third International Conference on Signal-Image Technology & Internet Based Systems (SITIS' 2007 ).* Shanghai, China, 2008.

[22] H. Sun, V. De Florio, N. Gui and C. Blondia. Towards Longer, Better, and More Active Lives - Building Mutual Assisted Living Community for Elder People. *In the Proceedings of the 47th European FITCE Congress, FITCE,* London 2008.

[23] T. Kleinberger, M. Becker, E. Ras, A. Holzinger, and P. Muller. Ambient Intelligence in Assisted Living: Enable Elderly People to Handle Future Interfaces, *Universal Access in Human-Computer Interaction. Ambient Interaction, Part II, HCII 2007.*

[24] M. Floeck, L. Litz, Aging in Place: Supporting Senior Citizens' Independence with Ambient Assistive Living Technology. *mst / news,* December 2007.

[25] H. Sun, V. De Florio and C. Blondia. A design tool to reason about Ambient Assisted Living Systems. *In the Proceedings of the International Conference on Intelligent Systems Design and Applications*, Jinan, China, 2006.

[26] H. Sun, V. De Florio, N. Gui and C. Blondia. Participant: A New Concept for Optimally Assisting the Elder People. *In the Proceedings of the 20th IEEE International Symposium on Computer-Based Medical Systems (CBMS-2007)*, Maribor, Slovenia, 2007.

[27] N. Gui, H. Sun, V. De Florio and C. Blondia. A Service-oriented Infrastructure Approach for Mutual Assistance Communities. *In proceedings of the IEEE Workshop on Adaptive and DependAble Mission- and bUsiness-critical mobile Systems (ADAMUS 2007),* Helsinki, Finland, 2007.

[28] Pekka Kahri, AAL-2009-2: “ICT based solutions ICT based solutions for Advancement of Social for Advancement of Social Interaction of Elderly People Interaction of Elderly People”, ICT 2008, Lyon.